# Conical intersections induced by the Renner effect in polyatomic molecules

Gabor J. Halász

*Department of Information Technology, University of Debrecen, H-4010 Debrecen P.O. Box 12, Hungary*

Ágnes Vibók

*Department of Theoretical Physics, University of Debrecen, H-40410 Debrecen P.O. Box 5, Hungary*

Roi Baer

*Department of Physical Chemistry, and the Fritz Haber Center for Molecular Dynamics, The Hebrew University of Jerusalem, 91904 Jerusalem Israel*

Michael Baer

*The Fritz Haber Center for Molecular Dynamics, The Hebrew University of Jerusalem, 91904 Jerusalem, Israel*

Characterizing and localizing electronic energy degeneracies is important for describing and controlling electronic energy flow in molecules. We show, using topological phase considerations that the Renner effect in polyatomic molecules with more than 3 nuclei is necessarily accompanied by "satellite" conical intersections. In these intersections the non-adiabatic coupling term is on the average half an integer. We present ab-inito results on the tetra-atomic radical cation $C_2H_2^+$ to demonstrate the theory.

The dynamics triggered in a molecule after absorbing a photon is usually discussed in terms of the Born-Oppenheimer theory[1], where the fast electronic degrees of freedom are treated separately from the slow nuclei. In this picture, electrons and nuclei do not easily exchange energy. Yet, in some nuclear configurations called conical intersections (CIs) energy exchange can become significant[1-6]. It is widely recognized today that were it not for CIs important photo-biochemical processes such as vision[7-9] and photosynthesis of vitamin-D[10] could not take place. CIs affect other important processes as well, such as photosynthesis in plants,[11] photochemistry of DNA[12], fluorescence of proteins[13, 14], molecular electronics[15, 16] and chemical dynamics[4, 17-19]. A major theoretical effort for the research of photochemistry and molecular light-harvesting processes is invested in the identification and localization of CIs[14, 19-25]. While numerical work is indispensable, there is still a fundamental need for understanding more on the ways CIs form. In few known cases CIs can be attributed to simple symmetry considerations[25]. Still, most CIs identified numerically seem to appear accidentally.

In this article, we report for the first time an unexpected connection between CIs and another type of electronic intersection, namely the Renner or Renner-Teller (R-T) intersection. We show, using their unique topological characteristics, that certain CIs must exist in a polyatomic molecule exhibiting the R-T effect[26] when these molecules are distorted from a linear configuration so that they lose both their axis and their plane of symmetry. These topological features can be revealed by considering a computationally accessible quantity, namely the line integral of the non-adiabatic coupling (NACT) vector along a closed contour in nuclear configuration space.

Within the Born-Oppenheimer treatment, one frequently has to consider simultaneously two-electronic adiabatic states because at some configurations they become degenerate. We denote these states as $\left|\zeta_j\left(\mathbf{s}_e\mid\mathbf{s}\right)\right\rangle$ $(j=1,2)$, where $\mathbf{s}_e$ and $\mathbf{s}$ are collections of electronic and nuclear coordinates, respectively. We will only consider real electronic wave functions (as long as spin orbit interactions are negligible this is permitted). Consider two such states which exhibit a CI. The corresponding electronic eigenvalues (also called adiabatic potential surfaces in Born Oppenheimer terminology), depend themselves on the nuclear coordinates. When the adiabatic potential surfaces are viewed as a function of two nuclear degrees of freedom (keeping all other fixed) they seem like two linearly intersecting cones, hence the name *ci*. Each adiabatic state changes sign when transported continuously along a closed loop enclosing the point of intersection[1]. It has been proved[1] that when this happens the line integral of the nonadiabatic coupling term (NACT) $\vec{\tau}_{12}(\mathbf{s})\left(\equiv\left\langle\zeta_1\mid\vec{\nabla}\zeta_2\right\rangle\right)$ along a closed contour $\Gamma$ equals:

$$\begin{aligned}
\alpha\left[\Gamma\right] &\equiv \oint_{\Gamma}\vec{\tau}_{12}(\mathbf{s})\cdot d\mathbf{s}\\
&= \pi\begin{cases}2n+1 & \Gamma \text{ encircles odd No. of cis}\\ 2n & \Gamma \text{ encircles even No. of cis}\end{cases}\\
n &= 0,\pm1,\pm2...
\end{aligned} \tag{1}$$

where $\alpha(\Gamma)$ is the corresponding topological (Berry) phase.

Eq. (1) reflects, among other things, the fact that the Curl condition[27], $\nabla\times\tau_{12}=0$, holds in the region surrounded by $\Gamma$ except at the *ci* points. In any actual calculation Eq. (1) holds only for a limited region in configuration space. For larger regions it is necessary to consider more than two adiabatic states and to apply the non-Abelian generalization of Eq. (1) and of the Curl condition[1] so that the conclusion are still valid. For simplicity we will not consider such an extension in this article.

A different type of intersection in the Born-Oppenheimer framework is the R-T intersection.[28-32] They occur when a molecule has a $C_\infty$ axis of symmetry i.e. the nuclei of the molecule are vibrating around a collinear configuration. We denote the direction of the collinear configuration as the $\hat{z}$ direction. In the exact collinear configuration the electronic angular momentum component $L_z=i\hbar\sum_n\dfrac{\partial}{\partial\theta_n}$ (where in the sum $n$ indexes all electrons) has the eigenvalue $\Lambda\hbar$ with integer $\Lambda\neq0$. Obviously, at collinearity $\Lambda$ is a good quantum number and the electronic energy must exhibit a two-fold degeneracy associated with $\pm\Lambda$. To be specific, let us consider the tetra-atom $C_2H_2^+$ cation radical, which is shown in the collinear configuration in Figure 1 (a). In collinearity, the electronic ground-state of this molecule is a doubly degenerate $\Pi$ state, namely, $|\Lambda|=1$.

Consider now a vibrational coordinate $q$ causing a small distortion out of the $C_\infty$ symmetry. For example, in the $C_2H_2^+$ one can take $q=q_1$ where $q_1$ is the displacement of the H1 atom out of the collinear configuration in Figure 1 (b). Once distorted, the symmetry reduces (to $C_s$). For small distortion, the two electronic eigenstates states $|\zeta_1\left(q\right)\rangle$ and $|\zeta_2\left(q\right)\rangle$ are predominantly just linear combinations of the two eigenstates at linearity $\left|\zeta_j\left(0\right)\right\rangle$. Since the eigenstates $\left|\zeta_j\left(q\right)\right\rangle$ are real, the electronic angular momentum component matrix element equals[28]

$$\langle \zeta_1(q) | L_z | \zeta_2(q) \rangle \approx i\hbar\Lambda \ (q \to 0) \tag{2}$$

where $\Lambda$ is an integer. Now, a rotation of all electrons by an angle $\varphi$ is equivalent to a rotation in the opposite direction of the nuclei, the associated NACT $\tau^{rigid}$ is equal to the angular momentum matrix element

$$\tau_{\varphi}^{rigid}(q \to 0) \equiv -\frac{i}{\hbar}\langle \zeta_1 | L_z | \zeta_2 \rangle = \langle \zeta_1 | \frac{\partial}{\partial\varphi} \zeta_2 \rangle = \tau_{\varphi12}(\varphi \mid q \to 0) = \Lambda \tag{3}$$

Next, after distorting the molecule by $q$ we perform a line integral, namely integrating over $\vec{\tau}_{12}$, obtaining (as long as q is not too large):

$$\alpha(\Gamma) = \oint_{\Gamma} \vec{\tau}(\mathbf{s}) \cdot d\mathbf{s} = \int_0^{2\pi} \tau_{\varphi}(q,\varphi)\,d\varphi \sim 2\pi$$
$$(\Gamma \text{ encircles the collinear R-T intersection axis}) \tag{4}$$

In what follows we extend the notation as follows: the angular NAC vector is now dependent on the position of the two hydrogen atoms $\tau_{\varphi_1}(q_1,\varphi_1;q_2,\varphi_2)$ is the NAC induced by motion of H1 and similarly $\tau_{\varphi_2}(q_1,\varphi_1;q_2,\varphi_2)$ refers to the NAC of H2.

In what follows we intend to show, first theoretically and then numerically, a closed loop in nuclear configuration space which has the following property: when the NAC vector is line-integrated along it, the result of the integration is $\pi$, thus revealing the existence of a CI. The loop we discuss is of type D in Figure 1. It is obtained by starting from a configuration where the two hydrogen atoms are displaced from the CC axis (such that all 4 atoms are on a plane) and then while holding atom H2 clamped the other atom, H1, is transported along a loop that surround the carbon-carbon bond. Note that this distortion (case D) is not a rigid rotation (the distance between H1 and H2 changes as the loop is transverse) and the path taken breaks not only the axial symmetry but also any planar symmetry of the distorted molecule.

To show that we consider first the loop in Figure 1(C) where both hydrogen atoms are shifted slightly out of collinearity (i.e. $q_1$ and $q_2$ are both different from zero) onto the same plane and then the plane is allowed to rotate around the carbon-carbon bond. This is a rigid rotation since all bond lengths do not change. Such a rotation in a Renner molecule will give a situation where the $\vec{\tau}_{12}$-NACT fulfills Eq. (3) so that the integration along a closed contour $\Gamma$ yields ~2$\pi$, as discussed in connection to Eq. (4). Writing the 2 dimensional line integral for this case we obtain the topological phase of path C in Figure 1:

$$\alpha(\Gamma_C) = \alpha(q_1,q_2) = \int_0^{2\pi} \big[ \tau_{\varphi_1}(q_1,\varphi;q_2,\varphi) + \tau_{\varphi_2}(q_1,\varphi;q_2,\varphi) \big]d\varphi = 2\pi \tag{5}$$

Note that mathematically equality is only obtained in the limit $q \to 0$, for finite but small values of $q$ the integral may in practice deviate by small amounts from $2\pi$. This will be seen in the numerical calculations we present bellow. Also, note that during this contour of integration the angle of both atoms H1 and H2 is equal and so there is only an integral on one angle $\varphi_1=\varphi_2=\varphi$.

But, now notice that the integral in Eq. (5) can be written also as the sum of two integrals, $\int_0^{2\pi} \tau_{\varphi_1}(q_1,\varphi;q_2,\varphi)\,d\varphi + \int_0^{2\pi} \tau_{\varphi_2}(q_1,\varphi;q_2,\varphi)\,d\varphi = 2\pi$. Thus, assuming the two rotating atoms to be identical (e.g. both are hydrogen atoms) and also $q_1 = q_2 = q$ we get that, due to symmetry, the two integrals are identical so that each contributes $\pi$ :

$$\int_0^{2\pi} \tau_{\varphi_1}\left(q, \varphi, q, \varphi\right) d\varphi = \pi \qquad (6)$$

Despite appearance Eq. (6) is not yet in the right form to say anything about the topological phase $\alpha\left(\Gamma_D\right)$ of path D in Figure 1. To do so, we need the line-integral formed by replacing the second $\varphi$ in the integrand by a fixed value, namely, $\varphi_{20}$:

$$\alpha(\Gamma_D) = \int_0^{2\pi} \tau_{\varphi_1}\left(q, \varphi, q, \varphi_{20}\right) d\varphi \qquad (7)$$

Assuming that the two atoms are far apart (in our case they are ~3.5 Å apart) and that q is small enough the replacement of $\varphi$ by $\varphi_{20}$ can only change the integral of Eq. (6) by a small amount, so $\alpha\left(\Gamma_D\right) \approx \pi$ . But we know that $\alpha$ , being a topological phase must be quantized to integer multiples of $\pi$ , we thus have exactly (in $q \rightarrow 0$ limit):

$$\alpha\left(\Gamma_D\right) = \pi \qquad (8)$$

This therefore means that we have proved that the path $D$ encircles a CI.

Before we discuss the meaning of this outcome we examine results of an *ab-initio* calculation along these lines of the tetra-atomic radical[33] $C_2H_2^+$. In these calculations, the C-C bond length is fixed at 1.254Å, and the z-axis component of the C-H bond length is fixed at 1.080Å. All calculations were made employing the MOLPRO program, using the state-average CASSCF level applying the 6-311G** basis set. The active space included all 9 valence electrons distributed on 10 orbitals. Five electronic states, including the states specifically studied, were computed by the state-average CASSCF level with equal weights. In certain cases these calculations were repeated with three to eight states to check for convergence. We first examine in Figure 2 (a), the values of $\tau_\varphi$(q) for two (schematic) cases given in Figure 1(b) (a single atom rotation) and 1(c) (rigid rotation of the two atoms). It is seen that in both cases, as long as q is small enough (<0.2 Å), the values of the two NACTs are $\Lambda \sim 1$ as asserted in Eq. (3). When q increases, the two-state approximation becomes less accurate (the system becomes non-Abelian) and the corresponding NACTs deviate from 1.

In Figure 2 (b) are presented the *ab-initio* calculated NACTs, $\tau_{\varphi_1}\left(q_1 = q, \varphi_1 = \varphi, q_2 = q, \varphi_2 = 0\right)$, as a function of $\varphi$ for three q-values, namely q=0.5, 0.3, 0.2 Å for the schematic case given in Fig. 1(d). The functions are seen to be slightly dependent on $\varphi$. It is noticed that on the average the curves approach the half-integer value, namely 0.5 as q decreases. Next, employing Eq. (7) for the curves in the figure, yields the *corresponding* values for the topological phases namely, $\beta$(q) = 3.01, 3.09, 3.12 Rads.

The results due to the above three situations are an indication that the line integral revealed the existence of a CI in the region surrounded by the corresponding circles. This formal connection between to radically different types of potential surface intersections is a new phenomenon. The fact the CI's must accompany the Renner effect may have important ramifications on the nonadiabatic dynamics in such molecules. To our knowledge, such an effect has never been taken into account and therefore needs to be examined in detail.

Following these findings the question is: where are the CI degeneracy points located. In the calculations we have performed we have seen that they are exceptionally close to the carbon-carbon axis (i.e. $q$ is very small). The few curves presented in Fig. 2(b) and the numerical values discussed subsequently support the expectation that these points are located *along* and *close* to the C-C axis (probably at an infinitesimal distance from that axis). It is noticed that as q decreases the above listed values (namely, 3.01, 3.09, 3.12 Rads.) $\rightarrow \pi$ (=3.14) which is an indication that the degeneracy points are indeed close to the C-C axis

In our opinion this study opens up a "Pandora box" for all kinds of topological effects taking place along the axes of collinear molecules as they contain more atoms.

We note that our proof has relied on the fact that the Renner states are of electronic angular momentum component equal to 1 (e.g. a pair of $\Pi$ state). The Renner effect exists also when the electronic states have even angular momentum (e.g. a pair of $\Delta$ states), but then our proof does not hold and we cannot guarantee the existence of CI's. Still, the fact that one type of intersection induces a second type of intersection raises an interesting question concerning the connection between the intrusive intersections breaking the Born-Oppenheimer treatment. More about this and other topological phenomena to be found for this system will be discussed elsewhere.

Support by the US-Israel bi-national Science Foundation, OTKA and NSF are acknowledged.

––––––––––––––––

Figure Captions:

**FIGURE 1: THE VARIOUS CONFIGURATIONS OF $C_2H_2^+$ DISCUSSED IN THE TEXT.**

**FIGURE 2: (A) THE NACT, $\tau_\varphi(q,\varphi)$, FOR THE SINGLE ATOM ROTATION (GIVEN IN FIGURE 1) AND FOR RIGID ROTATION GIVEN IN FIGURE 1 AS A FUNCTION OF Q: ooooo SINGLE ATOM; △△△△△ RIGID ROTATION. (B) THE NACT $\tau_{\varphi_1}(q_1 = q, \varphi, q_2 = q, \varphi_2 = 0)$ AS A FUNCTION OF φ FOR THREE DIFFERENT Q-VALUES: - - - - Q = 0.5 Å; ——— Q = 0.3 Å;; ●●●●● Q = 0.2 Å;. (SEE FIG. 1(D)).**

FIG 1

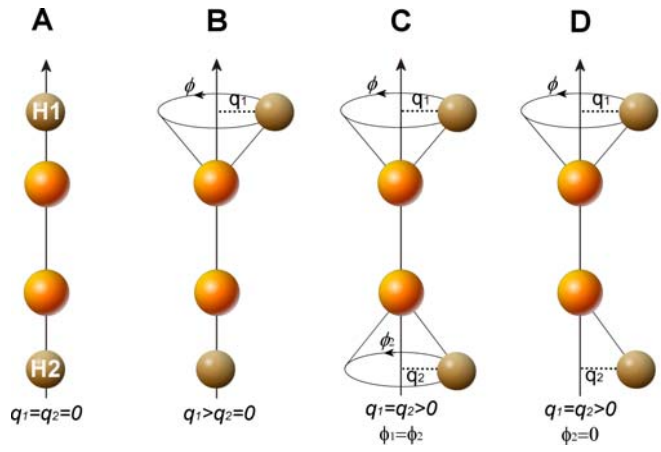

A
$q_1=q_2=0$

B
$q_1>q_2=0$

C
$q_1=q_2>0$
$\phi_1=\phi_2$

D
$q_1=q_2>0$
$\phi_2=0$

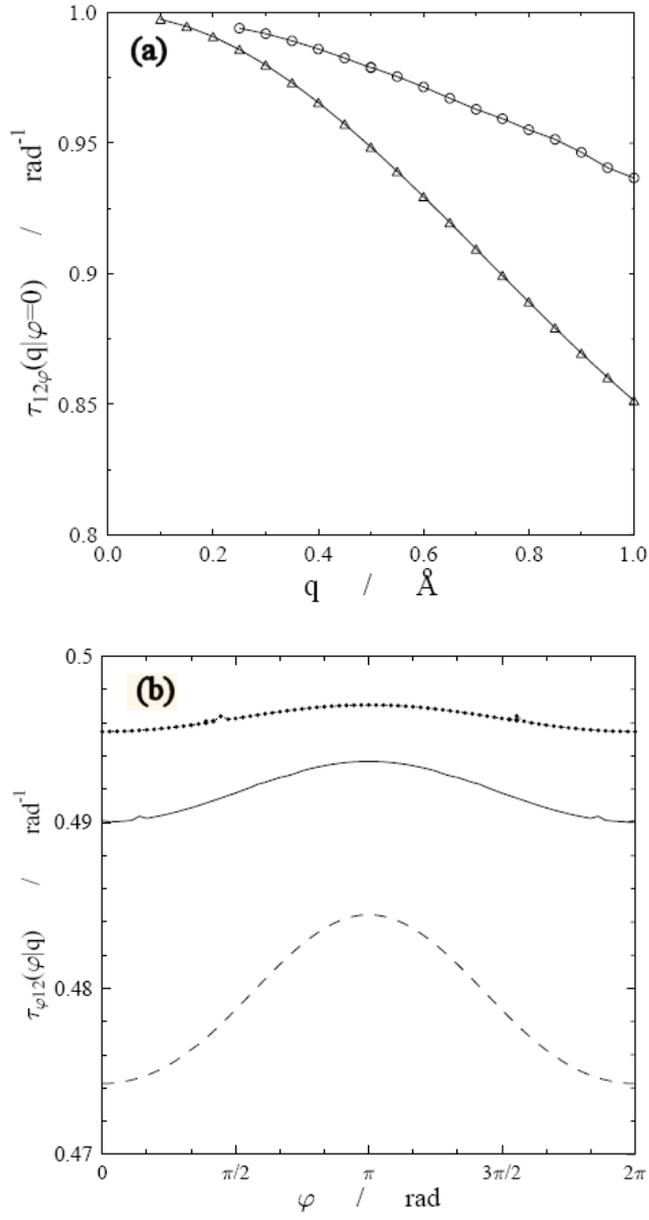